\documentstyle[psfig,graphicx,amssymb]{mn} 
\title[Cold gas in elliptical galaxies] {Cold gas in 
elliptical galaxies} 

\author[Georgakakis et al.] 
{A. Georgakakis$^{1}$\thanks{\sf age@star.sr.bham.ac.uk},
A. M. Hopkins$^{2}$,
A. Caulton$^{1}$,
T. Wiklind$^{3}$,
A. I. Terlevich$^{1}$,
\\\\
{\LARGE Duncan A. Forbes$^{4}$}
\\\\
  $^1$ School of Physics and Astronomy, University of Birmingham,
  Edgbaston, Birmingham, B15 2TT, UK\\
  $^2$  Department of Physics and Astronomy, University of Pittsburgh, 3941
  O'Hara Street, PA 15260, USA\\
  $^3$ Onsala Space Observatory, S-43992, Onsala, Sweden\\
  $^4$ Astrophysics \& Supercomputing, Swinburne University,
  Hawthorn, VIC 3122, Australia\\
}

\begin{document}
\maketitle  

\begin{abstract}
We explore the evolution of the cold gas (molecular and neutral
hydrogen) of elliptical galaxies and merger remnants ordered into a time
sequence on the basis of spectroscopic age estimates.
We find that the fraction of cold gas in early merger remnants
decreases significantly for $\approx$1--2\,Gyr, but subsequent evolution
toward evolved elliptical systems sees very little change. This trend can
be attributed to an initial gas depletion by strong star-formation which
subsequently declines to quiescent rates. This explanation is consistent
with the merger picture for the formation of elliptical galaxies.
We also explore the relation between H\,I--to--H$_2$ mass ratio
and spectroscopic galaxy age, but find no evidence for a statistically
significant trend. This suggests little net H\,I to H$_2$ conversion
for the systems in the present sample.
\end{abstract} 
 
\begin{keywords}
  Galaxies: mergers -- galaxies: starburst -- radio continuum: galaxies 
\end{keywords} 

\section{Introduction}

One of the proposed mechanisms for the formation of elliptical 
galaxies is the ``merger hypothesis" postulating that these systems 
are the product of disc galaxy mergers (Toomre \& Toomre 1972). 
A number of studies suggest that although merger remnants have many
properties that are different to those of ellipticals they will
eventually evolve to resemble these systems. Several aspects of this evolution
are briefly discussed by Georgakakis, Forbes \& Norris (2000; Paper I).
In particular, investigation of post-mergers at different evolutionary
stages suggests that global properties such as star-formation,
optical/far-infrared colours, and X-ray luminosity, evolve with time to
values typical of old ellipticals (Casoli et al. 1991;
Keel \& Wu 1995; Georgakakis, Forbes \& Norris
2000; O'Sullivan, Forbes \& Ponman 2000a).

The ``merger hypothesis" also faces some serious difficulties. One of the
long standing problems of this scenario is the excess globular clusters (GCs)
per unit starlight around ellipticals compared to spirals (van den Bergh
1984). Although evidence is accumulating that new GCs are formed during
interactions (Whitmore \& Schweizer 1995; Schweizer 1996; Miller et al. 1997;
Zepf et al. 1999), it is still unclear whether these objects will
evolve into {\em bona-fide} GCs (Brodie et al. 1998). Moreover, the
properties of the GC population of many old ellipticals, when
studied in detail, appear to be inconsistent with the merger scenario,
suggesting that not all ellipticals form by mergers (Forbes et al. 1997).

Another challenge for the merger picture comes from the properties of
cold gas in post-merger systems. In particular, the evolution of cold gas
(H\,I and H$_2$) was investigated in Paper I along an age sequence
comprising both pre- and post-mergers. A clear decrease can be seen in
cold gas content with age after the merger event, but even the most
advanced merger remnants in the sample (at ``age parameters" of
$\approx1.5\,$Gyr) are still richer in cold gas than evolved ellipticals
by about an order of magnitude. A number of mechanisms have been proposed
that may account for this discrepancy, including gas depletion by residual
star-formation activity, and gas heating to X-ray temperatures.
Evidence exists that suggests such processes do occur in merger remnants
(e.g. Hibbard \& van Gorkom 1996), but it remains unclear whether there
is a true evolutionary link between the cold gas properties of post-mergers
and ``normal" ellipticals. The merger remnants investigated in Paper I,
however, were limited to ``age parameters" (ages relative to the merger event)
in the range 0.1--1.5\,Gyr, and allow no conclusions to be drawn about
cold gas evolution at later stages. This paper investigates these questions
by extending the merger remnant sample of Paper I to include elliptical
galaxies spanning a much broader range of spectroscopic ages. New CO(1-0)
emission line data (estimating the H$_2$ gas mass) for merger remnants in
Paper I are also presented.

In Section~\ref{sample} we discuss the sample selection, while
Section~\ref{observations} describes new radio observations carried out for
some of the galaxies in the present sample. Section~\ref{results}
presents the results from our analysis. Finally, in
Section~\ref{conclusions} we summarise our conclusions. Throughout this
paper we assume a value $H_{0}=75\,\mathrm{km\,s^{-1}\,Mpc^{-1}}$.

\section{The Sample}\label{sample}

One of the difficulties in studying the evolution of ellipticals is
ordering them into a time sequence. We address this issue using a new 
catalogue of spectroscopic galaxy ages recently compiled by Terlevich
\& Forbes (2001) using a relatively homogeneous dataset of
high quality absorption line measurements for galaxies (e.g. H$\beta$
and [MgFe]). The models of Worthey (1994) are applied to this
dataset to break the age/metallicity degeneracy providing both
age and metallicity estimates for each galaxy. There are several
additional caveats regarding the reliability of spectroscopic ages.
The uniform assumption of solar element abundance ratios, in particular,
is a concern, as are effects from aperture sizes which sample different
physical scales, nebular H$\beta$ emission ``filling in" the age-sensitive
absorption feature, and others. These are all addressed in detail by
Terlevich \& Forbes (2001), who emphasise that although {\em absolute}
galaxy ages based on spectroscopic techniques are still under debate due to
theoretical and observational uncertainties, the ages quoted in their
catalogue provide a {\em relative} time-scale that can be reliably used
to order galaxies into an evolutionary sequence. For the purposes of
the present investigation the large catalogue of Terlevich \& Forbes (2001)
provides consistent, and to the extent allowed by the assumptions accurately,
derived age-estimates. This is a valuable tool for the purpose of defining
an evolutionary sequence, and, despite the concerns in the spectroscopic
age derivation, appears at least as effective as alternative methods,
and has thus been adopted here.

The present sample comprises all the elliptical galaxies from the
Terlevich \& Forbes (2001) catalogue with $M_{B}<-18.5$\,mag (total of
79). This absolute magnitude limit is adopted to avoid dwarf systems
(total of 10), likely to have different evolutionary histories
compared to more massive galaxies.

Additionally, the elliptical galaxy catalogue above is complemented
by a sample of 12 merger remnant candidates  compiled by Keel \& Wu
(1995) and comprising systems younger than $\approx1.5$\,Gyr. They used
morphological and dynamical criteria to assign a dynamical `stage'
number to each system estimating the time since nuclear
coalescence. The Keel \& Wu sequence is calibrated using spectroscopic
age estimates for some of the systems the sample (i.e. NGC\,2865,
NGC\,3921, NGC\,7252; Forbes, Ponman \& Brown 1998).    

The sample used in this study is shown in Table 1, which has the
following format:

{\bf  1.} Galaxy name.

{\bf  2.} Heliocentric distance, $D$, in Mpc, assuming $H_{o}= 75 \,
\mathrm{ km \, s^{-1} \, Mpc^{-1}}$.  No correction for the Local Group 
velocity or the Virgocentric flow has been applied. These corrections
are not expected to modify the estimated distances by more than $10\%$.
Moreover, in our analysis we consider ratios of observed
quantities that are independent of distance. 

{\bf 3.} Total radio flux density at 1.4\,GHz (20\,cm;
$S^{tot}_{1.4}$) in mJy. For most galaxies in the present sample
$S^{tot}_{1.4}$ was obtained from (i) Condon et al. (1991) (ii) the
NRAO VLA Sky Survey (NVSS) catalogue (Condon et al. 1998) and (iii)
the FIRST survey (Becker, White \& Helfand 1995).

{\bf 4.} Spectroscopic galaxy age in Gyr.

{\bf 5.} Far-infrared luminosity in solar units
($L_{\odot}=3.83\times10^{26}$\,W)

\begin{equation}\label{eq_lfir}
L_{FIR}=4\pi\,D^{2}\times 1.4 \times S_{FIR},
\end{equation}

\noindent where $S_{FIR}$ is the FIR flux in $\mathrm{W\,m^{-2}}$
between $42.5$ and $122.5\mu m$ (Sanders \& Mirabel 1996)

\begin{equation}\label{eq_fir} 
S_{FIR}(\mathrm{W\,m^{-2}})=1.26\times\,10^{-14}\times(2.58\times f_{60} + f_{100}), 
\end{equation}

\noindent where $f_{60}$ and $f_{100}$ are the IRAS fluxes at 60 and
100$\mu m$  respectively in Jansky. The scale factor 1.4 in equation
(\ref{eq_lfir}) is the correction factor required to account principally
for the extrapolated flux longward of the IRAS 100$\mu$m filter  (Sanders
\& Mirabel 1996).    

{\bf 6.} Molecular hydrogen mass, $M(H_2)$, estimated from the
CO(1--0) emission. The sources from which the CO(1--0) intensity measurements 
were obtained are given in Table 1. The conversion factor
$N(H_2)/I_{CO}=3\times10^{20} \, \mathrm{cm^{-2}\,(K\,km\,s^{-1})^{-1}}$,
appropriate for molecular clouds in the Milky Way (Sanders, Solomon \&
Scoville 1984) was adopted. It should be noted that use of this conversion
factor assumes  that the mean properties of the molecular gas in
ellipticals  and merger remnants (i.e. density, temperature and
metallicity) are similar to those  of the Milky Way.  This effect is
discussed in more detail later in the paper. Nevertheless, to
facilitate comparison of  our results with other studies we use the
standard Galactic $N(H_2)/I_{CO}$ conversion factor.  In any case the
results can be interpreted in terms of CO luminosity ($L_{CO}$) rather
than $M(H_2)$, since a constant scaling factor is used throughout. In
case of non-detections, upper limits were calculated by taking the
$3\sigma$ rms noise per channel multiplied with a standard line-width
of $300\,\mathrm{km\,s^{-1}}$ and divided by the number of channels in 
the adopted velocity interval.  

{\bf 7.} Neutral hydrogen mass, $M(HI)$. The H\,I masses are
related to the H\,I integrated intensities, $F(H\,I)$ (measured in 
$\mathrm{Jy\,km\,s^{-1}}$), by

\begin{equation}
M(H\,I) (M_{\odot})=2.356\times10^{5}\,\times F(H\,I)\,\times (D/\mathrm{Mpc})^{2}.
\end{equation}

\noindent The sources from which the H\,I intensity measurements were
obtained are also given in Table 1. Upper limits are calculated from
the equation above by taking the $3\sigma$ rms noise and assuming a
line-width of $300\,\mathrm{km\,s^{-1}}$.  Some ellipticals in our
sample have a late type spiral galaxy companion within the H\,I beam 
that dominates the observed H\,I emission. These systems (NGC 4105,
NGC 1209,  NGC 5638, NGC 7562) are excluded from the statistical
analysis of quantities involving H\,I gas mass and are shown on the
plots with different symbols.

{\bf 8.} Total $B$-band magnitude, $B_T$. This has been corrected for
Galactic extinction, $A_B$, but not for internal extinction. The  $A_B$
values are from Burstein \& Heiles (1984).

{\bf 9.} Galaxy X-ray luminosity, $L_X$, in units of
$\mathrm{erg\,s^{-1}}$ from O'Sullivan, Forbes \& Ponman  (2000b)
scaled to the galaxy distances adopted in the present study. 

{\bf 10.} Galaxy environment, field (F), group (G) or cluster (C)
from the  the catalogues of Tully (1988) and Garcia (1993).

\begin{figure*}
\centering
\vspace{-3.5cm}
\hspace*{-1cm}
\includegraphics[width=1.2\textwidth, height=1.2\textheight,angle=0]{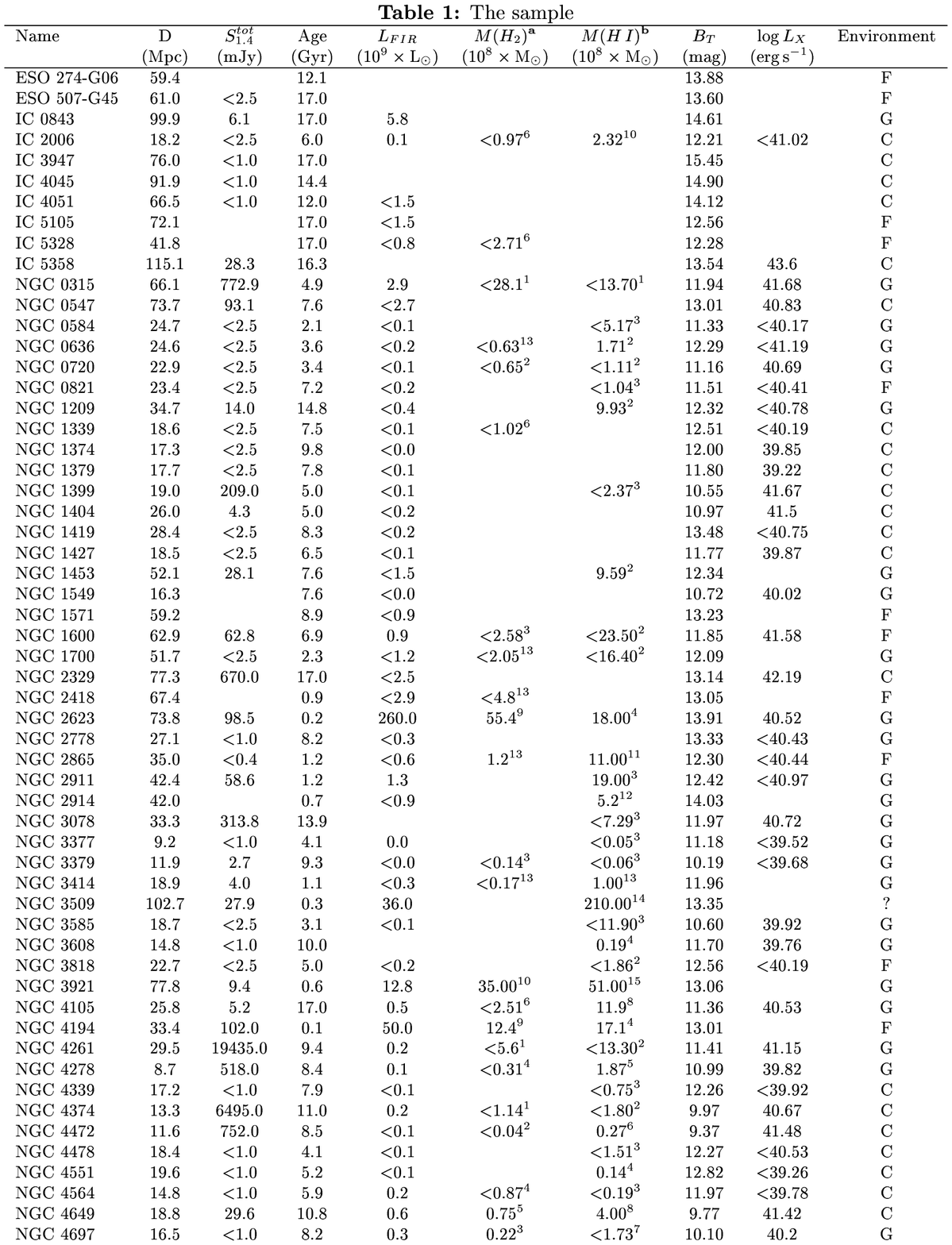}
\end{figure*}

\begin{figure*}
\centering
\vspace{-3.5cm}
\hspace*{-1cm}
\includegraphics[width=1.2\textwidth, height=1.2\textheight,angle=0]{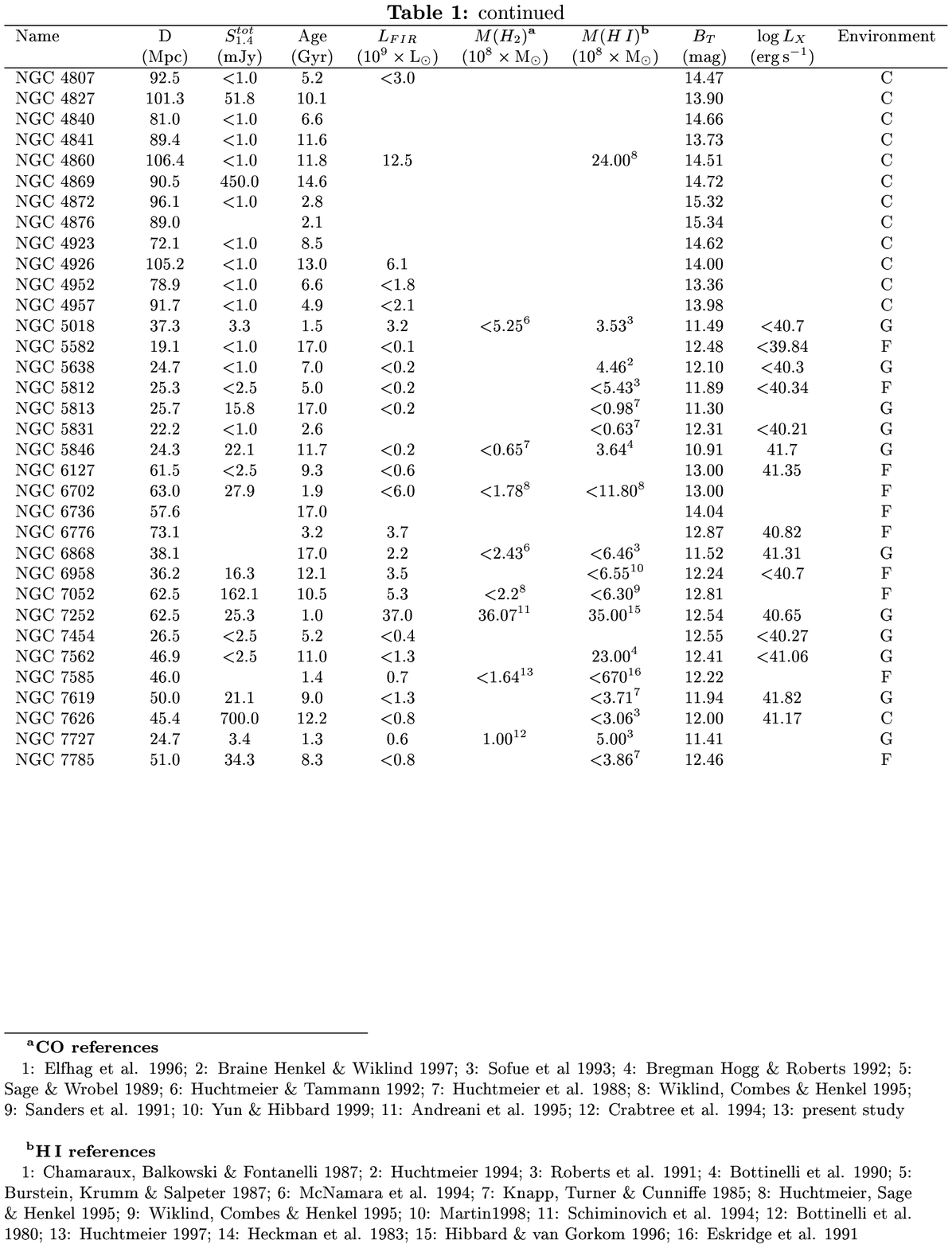}
\end{figure*}

\section{Observations}\label{observations}

New observations in the $^{12}$CO($J=1-0$) (115\,GHz) emission line 
were carried out in 2000 April using the Onsala Space Observatory
(OSO) 20\,m telescope. For low declination objects, additional 
observations in the $^{12}$CO($J=1-0$)  and $^{12}$CO($J=2-1$) (230\,GHz) 
emission lines were performed in 2000 October 10--13 at the Swedish--ESO 
Submillimeter Telescope (SEST). 

The OSO observations were obtained with a single-sideband SIS receiver
that operated at a typical system  temperature of 450-700\,K depending
on weather conditions and elevation. The beamwidth and main beam
efficiency of the OSO telescope at 115\,GHz were 33\,arcmin and 0.45
respectively. The spectra were obtained with a $512\times1$\,MHz
filterbank, yielding  a velocity resolution of
$2.6\,\mathrm{km\,s^{-1}}$. A dual beam switch mode with a beam-throw
of about 6\,arcmin was used to eliminate asymmetries and to produce
spectra with flat  baselines. 

For the SEST observations the SIS 115/230\,GHz  receiver was employed 
with two low resolution Acousto Optical Spectrometers (AOS) as backends. 
The AOS had a bandwidth of 500\,MHz  for the $J=1-0$  transition yielding 
a resolution of  about  $1.8\mathrm{km\,s^{-1}}$. For the $J=2-1$
transition the bandwidth was 1000\,MHz giving resolution of
$0.9\,\mathrm{km\,s^{-1}}$. Typical system temperatures were about  
300 and 200\,K for the $J=1-0$ and  $J=2-1$  transitions respectively.    
NGC\,2865 was observed in the CO(1--0) emission line only using the SIS
100\,GHz receiver because of technical problems with the SIS
115/230\,GHz receiver at the time of the  observation.  The backend was
again the  low resolution AOS with a bandwidth of 1000\,MHz yielding a
resolution of about $1.8\,\mathrm{km\,s^{-1}}$.  All observations were
made in dual beam switch mode with a beam-throw  of 11.2\,arcmin. The
half power beam sizes and main-beam efficiencies  of the SEST are
44\,arcsec and 0.70 for the $J=1-0$ transition and  23\,arcsec and
0.50 for the $J=2-1$ transition. 

The resulting spectra were reduced using the 
{\sc
xspec}\footnote{http://www.ls.eso.org/lasilla/Telescopes/SEST/handbook/}     
 software. Only first order baselines were subtracted and the
resultant spectra were ``boxcar'' smoothed to a resolution of 
$\approx10$ or $20\,\mathrm{km\,s^{-1}}$. Antenna temperatures,
$T_{A}^{*}$, were converted to main beam temperatures, $T_{mb}$, using
the relation
\begin{equation}
T_{mb}=T_A^{*}/\eta_{mb},
\end{equation}
where $\eta_{mb}$ is the main beam efficiency. The spectra of the
observed galaxies, in units of main beam temperature are shown in
Figure~1, while observational parameters are presented in Table 2. 
Only one confident detection has been measured, that for the CO(1--0)
transition in NGC\,2865. The integrated line intensity for this transition
$\int{T_{mb}\,dv}=0.55\,$K\,km\,s$^{-1}$. For the remaining galaxies
confident upper limits on $M(H_{2})$ can be derived from the non-detections.
These upper limits were calculated following the method described in
Section~\ref{sample}.   

For NGC\,6702 there appears to be a statistically significant
feature at a velocity which is $\approx250\mathrm{\,km\,s^{-1}}$ lower
than the optical systemic velocity. Wiklind, Combes \& Henkel (1995)
using data of similar sensitivity do not detect any  CO(1--0) emission
from this galaxy. If the feature in the NGC\,6702 is real it
corresponds to an intensity of $0.45\mathrm{K\,km\,s^{-1}}$ and a
molecular gas mass of $M(H_2)=3.20\times10^{8}\,M_\odot$. We adopt a
conservative approach and assume that no CO(1--0) is detected in this
system. More sensitive observations are required to confirm or refute the
observed feature.  

\begin{table*} 
\footnotesize 
\begin{center} 
\begin{tabular}{c ccc ccc c cc cc c}
\multicolumn{12}{l}{{\bf Table 2.} Observational parameters} \\ \hline
       & \multicolumn{3}{c}{}      & \multicolumn{3}{c}{}     &    
& \multicolumn{2}{c}{CO(1--0)}  & \multicolumn{2}{c}{CO(2-1)} & \\
 Galaxy& \multicolumn{3}{c}{RA}    & \multicolumn{3}{c}{DEC}  & $v^{(a)}$   
& $\delta T_{mb}^{(b)}$ & $\delta v^{(c)}$
& $\delta T_{mb}^{(b)}$ & $\delta v^{(c)}$ & Telescope\\
 Name  & \multicolumn{3}{c}{(J2000)} & \multicolumn{3}{c}{(J2000)} 
& ($\mathrm{km\,s^{-1}})$ & (mk)  
&  ($\mathrm{km\,s^{-1}})$ & (mk)  
&  ($\mathrm{km\,s^{-1}})$ \\ \hline

NGC\,0636  & 01 & 39 & 06.5 & -07 & 30 & 46 &  1847 &  3.2 & 10.9
& 3.6 & 5.5 & SEST\\
NGC\,1700 & 04 & 56 & 56.2 & -04 & 51 & 56 &  3915 &  2.5 & 10.9 &
2.6 & 5.5 & SEST \\
NGC\,2418 & 07 & 36 & 37.5 & 17 & 53 & 02.3 & 5057 &  4.6 & 21.2 &
- & - & OSO\\
NGC\,2865 & 09 & 23 & 30.1 & -23 & 09 & 43 &  2611 & 2.0 & 21.8
& - & - & SEST\\
NGC\,3414 & 10 & 51 & 16.3 & 27 & 58 & 28.4 & 1434 &  2.1 & 20.9 &
- & - & OSO \\
NGC\,6702  & 18 & 46 & 57.6 & 45 & 42 & 19.8 & 4712 &  4.8 & 20.8
& - & - & OSO \\
NGC\,7585 & 11 & 18 & 25.4 & 58 & 47 & 10.7 & 1787 &  2.8 & 10.9 & 3.3 & 5.5 & SEST \\
\hline
\multicolumn{12}{l}{(a) optical velocity}\\
\multicolumn{12}{l}{(b) Channel--to--channel noise rms}\\
\multicolumn{12}{l}{(c) Channel width used to derive rms and to plot
spectra} \\ 
\end{tabular}
\end{center} 
\end{table*}

\begin{figure*}
\centering
\includegraphics[width=0.9\textwidth, height=0.9\textheight,angle=0]{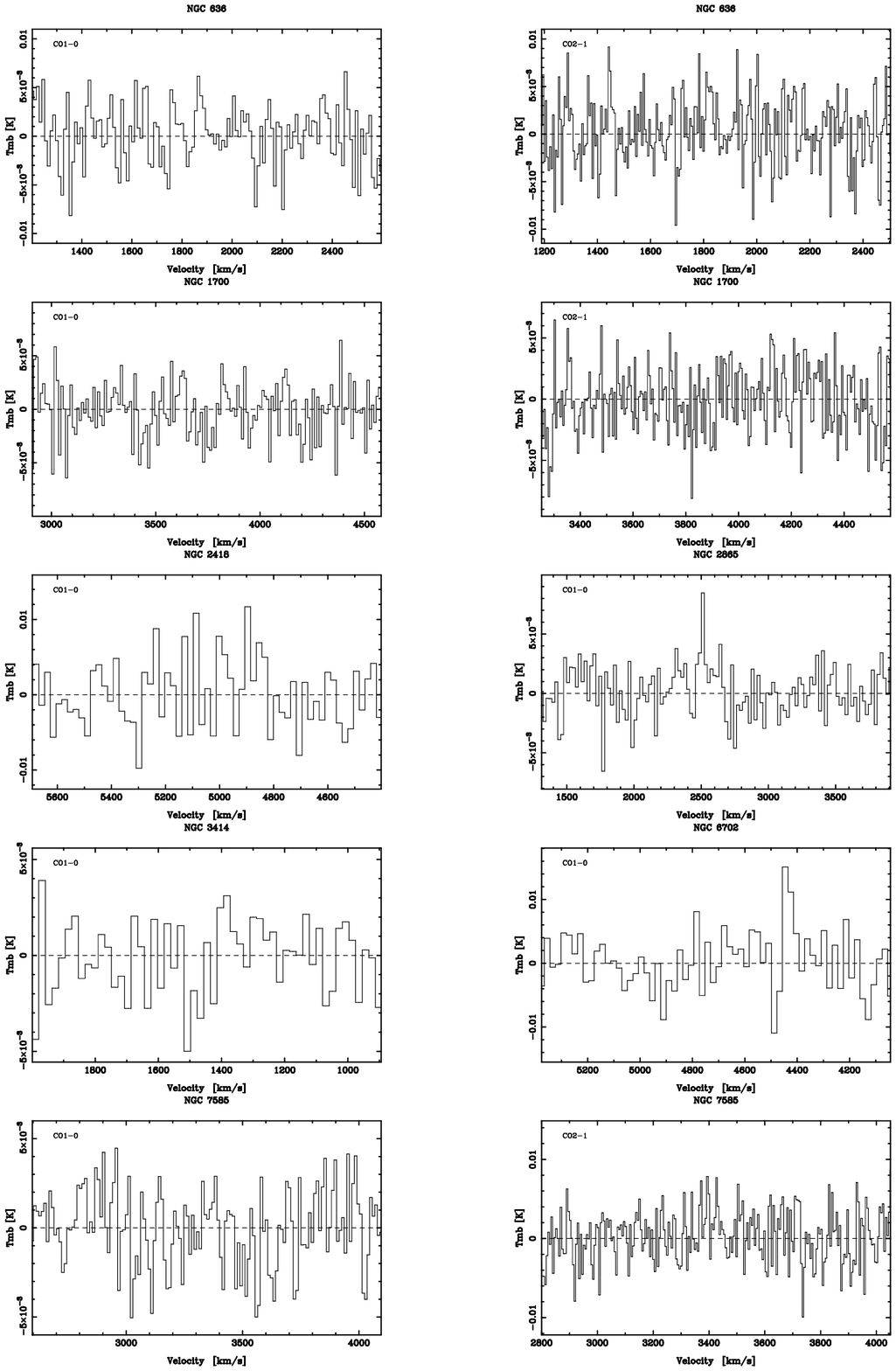}
\caption{Spectra of the observed galaxies. The channel-to-channel velocity resolutions
are the same as those given in Table 2. Galaxies NGC\,636, NGC\,1700, NGC\,2865 and
NGC\,7585 were observed with the SEST. NGC\,2418, NGC\,3414 and
NGC\,6702 were observed with the OSO 20\,m telescope}
\end{figure*}

\section{Results}\label{results}

\subsection{Star formation}\label{sec_sfr}

The star-formation activity in galaxies is believed to be tightly
associated with the presence of cold gas. Therefore, to better understand 
the evolution of cold gas in ellipticals, we first explore the
evolution of their star-formation activity. Here, we choose to use the
1.4\,GHz radio emission as estimator of the galaxy star-formation rate
(SFR). The same results and conclusions are obtained if the FIR luminosity is
used as a SFR census. It is important to recognise that the 1.4\,GHz flux
density of radio bright ($\approx1$\,Jy) ellipticals is dominated by powerful
active galactic nuclei (AGN) rather than star-formation. It is still unclear,
though, whether the low-level radio emission (at mJy and sub-mJy flux
densities) observed in many ellipticals is due to weak AGNs or residual
star-formation activity (Wrobel \& Heeschen 1991; Ho 1999;
Hopkins et al. 2000). Therefore, in the following analysis the radio
emission may need to be considered as an upper limit to the SFR of the
host galaxy.   

We define the ratio, $R$, between total radio (1.4\,GHz) flux density,
$S^{total}_{1.4}$ and $B$-band luminosity (Hummel 1981)

\begin{equation}\label{eq_r}
R=\log(S^{total}_{1.4})+0.4\times(B_{T}-12.5).
\end{equation}

\begin{figure} 
\centerline{\psfig{figure=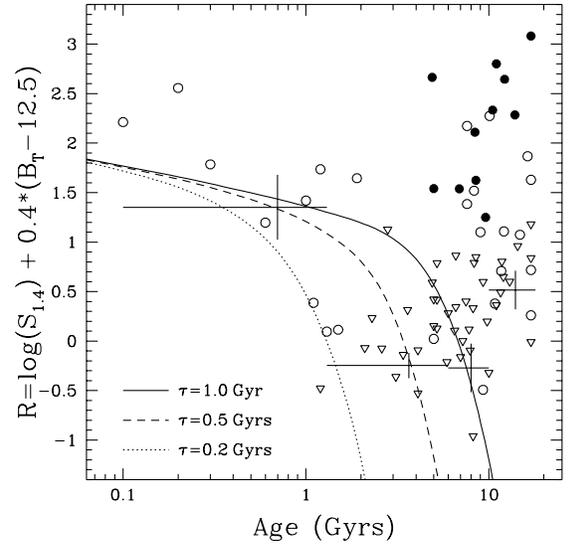,width=0.45\textwidth,angle=0}} 
\caption{Total radio to $B$-band flux ratio $(R=\log
S^{tot}_{1.4}+0.4\times[B_{T}-12.5])$ as a function of galaxy
age. Open circles are detections while triangles represent upper
limits. Filled circles are ellipticals with radio morphology
or optical spectroscopic features indicative of powerful AGN activity.
The crosses signify the mean $R$ for ellipticals averaged within age
bins. The horizontal error bars represent the width of the bin and the
vertical error bars are the standard error on the mean. The curves are
the models described in the text.
}\label{fig_r}
\end{figure}

\noindent The $R$ parameter is independent of distance and estimates the
ratio between radio power and optical luminosity. It has
been demonstrated that the mean radio power is proportional to the  mean
optical luminosity of galaxies (Hummel 1981). The $R$ parameter
also takes into account this effect, providing an estimate of the excess
radio emission in galaxies due to star-formation or AGN activity.  
The $R$ parameter is plotted against the galaxy age in Figure~\ref{fig_r}.

Galaxies with evidence for strong AGN activity based on either their
optical spectroscopic features (Veron-Cetty \& Veron 1998) or their
radio morphology (e.g. radio lobes; if high resolution radio maps are
available) are excluded from subsequent statistical analysis. This ensures
that the trends investigated more reliably reflect the evolution of
star-formation processes. This piecemeal method for excluding AGN systems
is not completely reliable, and a number of the remaining old ($>3$\,Gyr)
ellipticals in the sample have comparatively high radio emission ($R>1.5$).
While the young ($<1$\,Gyr) post-merger systems with similarly high
$R$ values show strong evidence for the presence of starburst activity
(Keel \& Wu 1995), this is much less typical of evolved elliptical galaxies.
The majority of these older systems have radio flux densities at the
1\,Jy level and are likely to belong to the ``classic'' AGN radio
population, although in the absence of unambiguous spectroscopic or
morphological classification, they have been retained in the
following analysis.

With these caveats in mind, and despite the many non-detections, there is
evidence for a decline in the star-formation rates (traced by radio
emission) from early merger-remnants to old ellipticals. This can
be demonstrated by estimating the mean
$R$ parameter within different age bins using survival analysis techniques
implemented in the {\sc asurv} package (LaValley, Isobe \& Feigelson 1992;
Isobe, Feigelson \& Nelson 1986). Because of the many upper limits, and
the probable (albeit unintentional) inclusion of several AGN systems, the
mean $R$ values in the last three bins are likely to be overestimates
of the level of star-formation present. This reinforces the suggestion
that the intensity of star-formation processes decreases with galaxy age.

Also shown in Figure~\ref{fig_r} are simple models that assume   
that the galaxy SFR follows an exponential decay of the form

\begin{equation}\label{eq_1}
SFR\propto e^{-t/\tau},
\end{equation}

\noindent 
where $t$ is the time since the onset of the star-formation and $\tau$
is the e-folding parameter. For this SFR law the population synthesis
code of Bruzual \& Charlot (1993) with a Salpeter IMF is used to
predict the galaxy $B$-band luminosity evolution. Additionally, to
estimate the 1.4\,GHz radio  luminosity ($L_{1.4}$), we assume 
that  $L_{1.4}$ is directly proportional to the  model galaxy
SFR. Following Condon (1992) and assuming a Salpeter IMF we find    

\begin{equation}
L_{1.4}=SFR\,\times\,8.6\times10^{20}\,\mathrm{(W\,Hz^{-1})},
\end{equation}

\noindent 
where the SFR is estimated from equation \ref{eq_1}. Models with
e-folding parameter $\tau\approx2\times10^8-1\times10^9$\,yrs can
reproduce the observed range of radio emission, with the exception of
powerful radio ellipticals, dominated by AGNs. It should be noted that
our aim is not to find the best fit to the  observed trend but to
demonstrate that simple models  assuming a burst of star-formation
that declines with time can  reproduce both the observed trend and the
the range of $R$ parameters from early merger-remnants to evolved
ellipticals. Indeed, the adopted model does not include important
processes such as recycling and heating of the gas. Additionally,  the
radio emission of some ellipticals in the present sample are likely to
be dominated by AGNs rather than SFR. Consequently, the adopted model
with $\tau\approx2\times10^8-1\times10^9$\,yrs is likely to represent
a lower envelope. Moreover, the spectroscopic ages  for the galaxies
in the present sample assume an instantaneous burst of star-formation 
rather than a continuous decline. Estimating spectroscopic ages using
models with finite bursts of star-formation will result in older
galaxy ages than the ones estimated here. The SFR in the Bruzual
\& Charlot models, however, declines fast enough that any differences
should be small for old systems (e.g. $>3-4$\,Gyr).

\begin{figure} 
\centerline{\psfig{figure=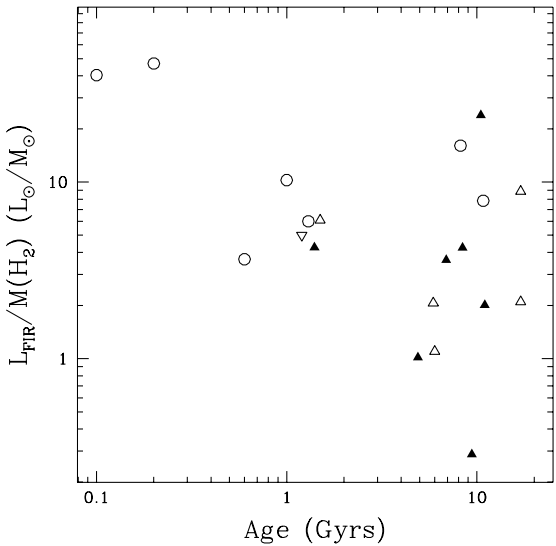,width=0.45\textwidth,angle=0}} 
\caption{Star-formation efficiency ($L_{FIR}/M(H_2)$) as a function of
galaxy age. Open circles are detections while triangles represent upper
and lower limits. Filled symbols are ellipticals with radio morphology
or optical spectroscopic features indicative of powerful AGN activity.
}\label{fig_sfe}
\end{figure}

We further explore the star-formation activity in ellipticals using 
the ratio of FIR luminosity to molecular hydrogen mass
$L_{FIR}/M(H_2)$. This estimates the number of massive stars 
formed per molecular cloud and is thus, related to the integrated 
galaxy star-formation efficiency (SFE).  Figure~\ref{fig_sfe} plots 
the SFE as a function of the age for the galaxies in the
present sample. There is significant scatter, especially for old
systems, and many lower limits due to uncertain $M(H_2)$
estimates. Excluding AGNs, the Kendal tau test gives a probability of
$78\%$ that no correlation is present, suggesting that the SFE of
ellipticals and merger remnants is roughly constant with time. There
is only tenuous evidence for a decline in the case of early merger
remnants and young ellipticals for ages $<2$\,Gyr.     

Therefore, although we find evidence that the SFR of ellipticals is
declining the efficiency of  the star-formation appears to remain
unchanged with age, with the possible exception of very young merger
remnants. This is in agreement with studies of the SFE variations 
among early type galaxies (Lees et al. 1991). It has been suggested,
however, that the FIR luminosity of 
evolved ellipticals originates from cold rather than warm dust, and
thus it is not a good estimator of the low-level SFR in these systems
(Lees et  al. 1991). Similarly, it has been proposed that the FIR
luminosity of ellipticals is not sensitive to on-going star-formation
but depends on both the mass-loss rate of evolved stars and the mass
of the hot X-ray emitting gas (e.g. Bregman, Hogg \& Roberts 1992). In
any case, more data, both FIR luminosities and $M(H_2)$  estimates,
are required to further  constrain the SFE of elliptical
galaxies.     

\subsection{Cold gas evolution}

\begin{figure} 
\centerline{\psfig{figure=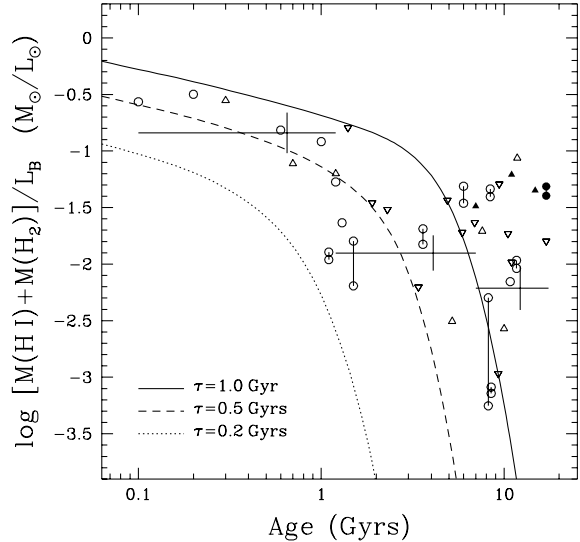,width=0.45\textwidth,angle=0}}
\caption{ Total mass of neutral and molecular hydrogen normalised to
the $B$-band luminosity as a function of galaxy age. Circles are
detections while triangles represent upper or lower limits. Points 
connected with a line represent the upper and lower
$[M(H_2)+M(HI)]/L_{B}$ limits for that system. Filled symbols are for 
ellipticals with nearby late type galaxy companions that dominate
the observed H\,I emission (NGC 4105, NGC 1209,  NGC 5638, NGC
7562; section 2), which are excluded from the analysis. The crosses
signify the mean cold gas content for
ellipticals averaged within age bins. The horizontal error bars
represent the width of the bin and the vertical error bars are the
standard error on the mean.
The curves are the models described in the text.  There is evidence
for a decrease in  $[M(H_2)+M(HI)]/L_{B}$ from early merger-remnants
to old ellipticals indicating cold gas depletion.
}\label{fig_mt}
\end{figure}

The ratio of cold gas mass (molecular and neutral hydrogen) to
$B$-band luminosity ($[M(HI)+M(H_2)]/L_{B}$) is plotted against
spectroscopic galaxy age in Figure~\ref{fig_mt}. Assuming that $L_B$
provides, to a first approximation, a measure of the total galaxy
mass, the  ratio $[M(HI)+M(H_2)]/L_{B}$ estimates the fraction 
of cold gas mass in the system. It is clear that this quantity
decreases from early merger remnants to young ellipticals on a time
scale of 1--2\,Gyr, as shown in Paper I. This is supported by the
generalised Kendal tau and the Spearman rho tests, which indicate a
strong anticorrelation between age and cold gas fraction at
a significance level of $99.99\%$. This result is dominated by
the gas rich young merger remnants. Applying these statistical
tests only to the sub-sample of old ($>1.5$\,Gyr) ellipticals,
we find weak or no correlations between cold gas content and
spectroscopic age. This suggests little further decline in cold
gas fraction after the initial sharp decrease.

In the context of the merger scenario for the formation of ellipticals,
this result is consistent with major changes in the cold gas content
of these systems due to depletion by star-formation, for example,
occuring predominantly by 1--2\,Gyr after the merger event, with a roughly
constant fraction of cold gas remaining during subsequent evolution.
If there is quiescent low-level star-formation, which would be expected
to continue to steadily reduce the cold-gas fraction, the latter effect
might be explained as a ``steady-state" scenario, either due to
(i) injection of cold gas from mass shed by evolved stars
(Faber \&  Gallagher 1976), (ii) fading
of the stellar population with age resulting in a decrease in $L_B$, 
(iii) the return of the H\,I expelled into tidal tails during early
stages of the interaction (Hibbard \& van Gorkom 1996). Alternatively,
it is possible that the mechanism responsible for the initial depletion
of the gas (e.g. SFR) declines
rapidly with time and hence, does not have any significant effect on
the remaining cold gas reservoirs for ages $>$1--2\,Gyr. Nevertheless,
despite the scatter in Figure~\ref{fig_mt}, there may be a hint
of a slow decline in cold gas fraction for evolved systems suggested
by the decreasing mean within the age bins shown, although this
effect is small enough that it may simply be an artifact of the
scatter.

It should be noted that the  present study assumes a standard
CO--to--H$_2$ conversion factor,
$N(H_2)/I_{CO}=3\times10^{20}\,\mathrm{cm^{-2}\,(K\,km\,s^{-1})^{-1}}$,  
appropriate for molecular clouds in the Milky Way  (Sanders, Solomon 
\& Scoville 1984). The molecular clouds in ellipticals and merger
remnants, though, might have properties (i.e. density and
temperature) different to those of the Milky Way. Such differences are
expected to modify the CO--to--H$_2$ conversion factor. Therefore, it is
possible that the declining trend between merger remnants and evolved
ellipticals in Figure~\ref{fig_mt} might be due to systematic
variations in the $N(H_2)/I_{CO}$ factor between young  merger
remnants and evolved ellipticals.  In particular, the  $N(H_2)/I_{CO}$
factor has been shown to depend on the mean 
temperature ($<T>$) and density ($<\rho>$) of the  intergalactic
medium according to the relation (Maloney \& Black 1988) 

\begin{equation}\label{eq_con}
N(H_2)/I_{CO} \propto <\rho>^{1/2}\, <T>^{-1}.
\end{equation}

\noindent It is clear that if the mean molecular gas temperature in
evolved ellipticals is lower than that in early merger remnants, due
to lower levels of star-formation heating up the clouds, then use of
the same conversion factor will systematically underestimate $M(H_2)$
in old ellipticals.  
On the contrary, the higher molecular gas densities expected in merger
remnants due to compression have the opposite effect on the
conversion factor. 
Therefore, if temperature variations are to explain the observed
trend in Figure~\ref{fig_mt}, large differences, by as much as
1.5\,dex, are required.  
Moreover, high molecular gas temperatures are usually associated with
enhanced H$_2$ mean densities. Consequently, variations of the
$N(H_2)/I_{CO}$ factor alone cannot easily explain the observed trend
in Figure~\ref{fig_mt}. 

The CO--to--H$_2$ conversion factor is also sensitive to the  
metallicity, $Z$,  of the galaxy. In particular, Arnault et al. (1998)
suggest that the $N(H_2)/I_{CO}$ varies with $Z^{-2.2}$. Therefore, if
metallicity variations are to explain the trend in Figure~\ref{fig_mt}
(1.5\,dex decline) then young merger remnants should be as much as
$\approx5$ times more metal rich than evolved ellipticals. Indeed,
a higher than solar metallicity translates to a lower $N(H_2)/I_{CO}$
factor compared to the Galactic one and hence, $M(H_2)$ is
overestimated when using the standard Galactic conversion
factor. Such a large metallicity excess from merger remnants to
evolved ellipticals would be difficult to explain.  
 
Also shown in Figure~\ref{fig_mt} are the predictions of the simple
model described in the previous section. In this model the cold gas
mass, $M_g$, is assumed to be the mass fraction of the galaxy that
remains to be used for star-formation (i.e. the mass of the galaxy not
in stars). It can be shown that for an exponential SFR the gas mass 
decreases  according to the relation

\begin{equation}\label{eq_2}
M_{g}\propto e^{-t/\tau}.
\end{equation}

\noindent 
This model with the {\em same} parameters as
used in the previous section, $\tau\approx2\times10^8 -
1\times10^9$, reproduces the observed trend from early merger remnants
to evolved ellipticals reasonably well, at least for ages
$<7$\,Gyr. On the contrary, older systems are gas rich compared to the
model prediction. This suggests that cold gas depletion due to on-going
star-formation is, at least partially, a plausible mechanism for the
observed trend in Figure~\ref{fig_mt}.

Previous studies of the cold gas content of merger remnants show that
these systems are gas rich compared to evolved ellipticals and S0s
(Hibbard et al. 1994; Paper I). To resolve this problem Hibbard
et al. (1994) suggested that low level residual SFR could further
deplete the gas reservoirs of these systems within few Gyr to values
typical to E/S0s. Clearly the trend in Figure~\ref{fig_mt} is
consistent with this scenario.  

\begin{figure} 
\centerline{\psfig{figure=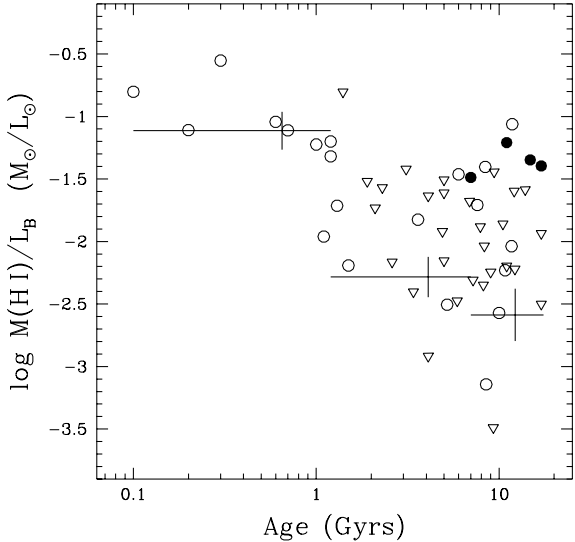,width=0.45\textwidth,angle=0}}
\caption{ Mass of neutral hydrogen normalised to $B$-band luminosity
as a function of galaxy age. Circles are detections while triangles
represent upper or lower limits. Filled symbols are for 
ellipticals with nearby late type galaxy companions that dominate
the observed H\,I emission (NGC 4105, NGC 1209,  NGC 5638, NGC
7562; section 2), which are excluded from the analysis. The crosses
signify the mean H\,I content for ellipticals averaged within age bins.
The horizontal error bars represent the width of the bin and the vertical
error bars are the standard error on the mean. There is evidence for
a decrease in $M(HI)/L_{B}$ from early merger-remnants to old
ellipticals indicating gas depletion. 
}\label{fig_mhi}
\end{figure}

Figure~\ref{fig_mhi} plots the H\,I mass fraction as a
function of spectroscopic galaxy age. As with the cold gas fraction
in Figure~\ref{fig_mt}, it is clear that $M(\mathrm{H\,I})/L_B$
also decreases, on average, from young merger remnants to evolved
ellipticals. This supports our conclusions drawn from the trends explored
in Figure~\ref{fig_mt}, suggesting they are unlikely to be artifacts of
the adopted CO--to--H$_2$ conversion factor. For Figure~\ref{fig_mhi}
the generalised Kendal tau and the Spearman rho tests again indicate a
strong anticorrelation between age and H\,I gas fraction at a
significance level of $99.99\%$, a result dominated by the young merger
remnants. For older systems ($>1.5\,$Gyr), there is again no statistically
significant trend with age, although the decline in the mean from the
second bin to the third is still suggestive.

Sansom et al. (2000) also studied the $M(H\,I)/L_B$ evolution of
ellipticals using the fine-structure index, $\Sigma$, to quantify
morphological peculiarities in the light profile of the galaxies and to
order them into an evolutionary sequence. They find no evidence of
a decrease in the cold gas along their evolutionary sequence for their
sample, which includes only 2 merger remnants and is dominated by
older systems ($>1.5$\,Gyr). This is consistent with the trend seen in
our sample over this age range.

\begin{figure} 
\centerline{\psfig{figure=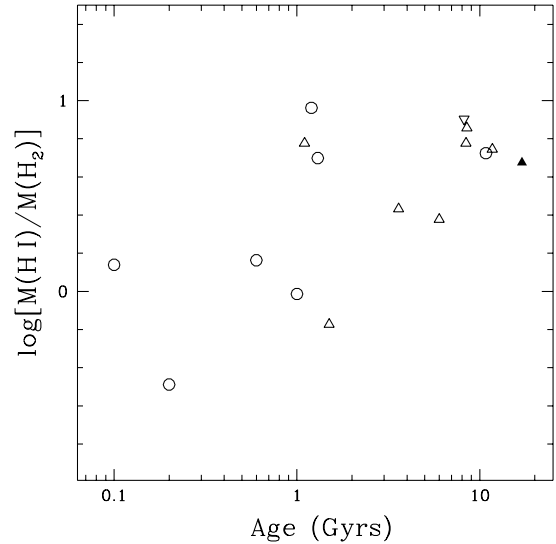,width=0.45\textwidth,angle=0}} 
\caption{Neutral to molecular hydrogen mass, $M(HI)/M(H_2)$, as a
function of galaxy age.  Filled symbols are for  elliptical
galaxies with nearby late type galaxy companions that dominate the
observed H\,I emission (section 2), which are excluded from the
analysis. There is no obvious trend,
implying little net conversion from H\,I to H$_{2}$ during the
interaction.  
}\label{fig_mhimh2}
\end{figure}

Finally, we explore whether net conversion of atomic to molecular
hydrogen is occurring in ellipticals. In particular, in the merger
picture for elliptical galaxy formation, the gravitational
instabilities  experienced during the interaction can force about half
of the  outer disc H\,I  into a tail, the rest of the H\,I being
forced into the inner regions (Hibbard \& Mihos 1995).
Hibbard \& Van Gorkom (1996), however, found little evidence for neutral
hydrogen within remnant bodies, with most of it lying in the outer
regions (i.e. tidal features). One of the proposed mechanisms to
resolve this discrepancy is  net conversion from H\,I to H$_2$.
Figure~\ref{fig_mhimh2}  plots the ratio of neutral to molecular
hydrogen mass as a function of spectroscopic galaxy age. Despite the
limited data available there is some evidence for a correlation,
suggesting that  net H\,I to H$_2$ conversion might be taking
place. The generalised Kendall tau test indicates a correlation,
albeit at the 2.5$\sigma$ significance level. This is primarily driven by
the NGC\,2623 which has the lowest H\,I/H$_2$ ratio in the
sample. Excluding this galaxy from the analysis, 
the Kendall tau test finds no significant correlation between age and
H\,I/H$_2$ ratio (probability of a correlation
$\approx70\%$). Therefore, we conclude that any net H\,I to H$_2$
conversion is small. 
Similarly,  searches for molecular hydrogen in the
merger remnants NGC\,7252 and NGC\,3921 have revealed that these
systems have below average molecular gas content compared with their
spiral progenitors (Solomon  \& Sage 1988; Young \&  Knezek 1989;
Hibbard \& van Gorkom 1996). This suggests that any net conversion of
atomic to molecular hydrogen is  relatively inefficient, in agreement
with our results.

\section{Conclusions}\label{conclusions}

In this paper we study the cold gas evolution of early
type galaxies and merger remnants. Spectroscopic ages, estimating the
time since the last major starburst event, are used to order the
galaxies in the present sample into an evolutionary sequence.      
We find strong evidence for a decrease of the cold gas mass
fraction (H\,I and H$_2$) from early merger remnants to evolved
ellipticals. The observed cold gas decline is predominantly occuring
1--2\,Gyr after the merger event, and the cold gas fraction remains
essentially constant thereafter. Simple models involving on-going
but declining star-formation activity provide a plausible explanation
for the observed trend.
This is consistent with the ``merger hypothesis", where the
interaction induced starburst is depleting the cold gas reservoirs of
the system to values typical for those of evolved ellipticals.
We also find little evidence for net conversion of H\,I to 
H$_2$ in merger remnants and ellipticals, particularly for ages $>1$\,Gyr,
in agreement with previous studies.

\section{Acknowledgements}
We wish to thank an anonymous referee for several very constructive and
helpful comments. AMH acknowledges support from NASA LTSA grant
NRA-98-03-LTSA-039. This research has made use of the {\sc nasa/ipac}
Extragalactic Database ({\sc ned}), which is operated by the Jet Propulsion
Laboratory, Caltech, under contract with the National Aeronautics and
Space Administration.

\end{document}